\documentclass[]{aa}
\usepackage{txfonts}
\usepackage{graphicx,epsfig,fancyhdr,rotating,amsmath,natbib}
\usepackage{color}

\begin{document}

   \title{Accretion in the recurrent nova T CrB: 
          Linking the superactive state to the predicted outburst \thanks{The data are  available on zenodo.org/records/10283430}} 
   \titlerunning{The recurrent nova T CrB} 
   \author{R. Zamanov\inst{1}\fnmsep\thanks{\email{rkz@astro.bas.bg}}
          \and   S. Boeva\inst{1}
          \and   G. Y. Latev\inst{1}
          \and   E. Semkov\inst{1}
          \and   M. Minev\inst{1}
          \and   A. Kostov\inst{1} 
          \and   M. F. Bode\inst{2,3}
          \and   V. Marchev\inst{1} 
          \and   D. Marchev\inst{4} 
          }
   \institute{Institute of Astronomy and National Astronomical Observatory, Bulgarian Academy of Sciences, 
              Tsarigradsko Shose 72, BG-1784 Sofia, Bulgaria  
         \and
      Astrophysics Research Institute, Liverpool John Moores University, IC2, 149 Brownlow Hill, Liverpool, L3 5RF, UK
         \and
      Office of the Vice Chancellor, Botswana International University of Science and Technology, 
      Private Bag 16, Palapye, Botswana         
        \and
       Department of Physics and Astronomy, Shumen University "Episkop Konstantin Preslavski",  
       115 Universitetska Str., 9700 Shumen, Bulgaria
             }

   \date{Received October 26, 2023; accepted December 5, 2023}

 
\abstract 
{ T CrB (NOVA CrB 1946) is a famous recurrent nova with a recurrence timescale of 80 years. }
{ We aim to estimate the colours, luminosity, and mass-accretion rate for T CrB (NOVA CrB 1946) during and after the superactive state. }
{ We performed and analysed $UBV$ photometry of the recurrent nova T~CrB. }
{ For the  hot component of T~CrB, we find average dereddened colours of
$(U-B)_0 = -0.70 \pm 0.08$ and $(B-V)_0 = 0.23 \pm 0.06$, which correspond to 
an effective temperature of $9400 \pm 500$~K and an optical luminosity of
$40-110~L_\odot$ during the superactive state (2016-2022).  
After the end of the superactive state, the hot component became 
significantly  redder, $(U-B)_0 \approx -0.3$ and  $(B-V)_0 \approx 0.6$ in August 2023,
and its luminosity decreased markedly to $20-25$~$L_\odot$ in April-May 2023, 
and to $8-9~L_\odot$ in August 2023.  
The total mass accreted during the superactive state 
from 2014 to 2023 is $\sim 2 \times 10^{-7}$~M$_\odot$.  }  
{This is a significant fraction of the mass required to cause a thermonuclear runaway (TNR). Overall our results support a model in which a large accretion disc acts as a reservoir with increased accretion rate onto the central white dwarf during disc high states, ultimately 
leading to a TNR explosion, which now seems to be imminent.}


   \maketitle
%

\section{Introduction}

T Coronae Borealis (HD 143454) is a famous recurrent nova, with recorded eruptions in 1866 and 1946
and possibly in 1217 and  1787 \citep{2023arXiv230813668S}.
A new outburst is expected over the coming 12 months \citep{2023MNRAS.524.3146S, 2023arXiv230810011M} and is the subject of a great deal of preparatory work 
among the international astronomical community. The consensus is that T CrB would be the brightest nova outburst observed since  Nova 1500 Cyg in 1975.
The likely presence of a dense wind from the red giant 
(see below) will also lead to shock systems 
as the outburst ejecta encounter it, 
as in the recurrent nova RS Oph \citep[e.g.][]{1985MNRAS.217..205B}. 
Major advances in understanding its  nature were made  
when \cite{1949ApJ...109...81S} 
discovered that the radial velocity of the M giant 
of T CrB varies with a period of 230.5 days 
and when \cite{1992ApJ...393..289S}, 
using $IUE$ spectra, identified that the hot component of the system is a white dwarf accretor.
In this binary, the ellipsoid-shaped red giant fills its Roche lobe and transfers material
via the inner Lagrangian point L$_1$.
This type of nova is also referred to as a symbiotic recurrent nova  
\citep[e.g.][]{2010AN....331..160B, 2011A&A...527A..98S}. 
Indeed, T~CrB shares characteristics with three types of interacting
binary (recurrent novae, symbiotic stars, and cataclysmic variables), 
and is an important object for understanding 
accretion, disc instabilities, and nova outbursts.  


From 2016 until 2023, T~CrB was in a superactive state \citep{2016NewA...47....7M,   
 2023RNAAS...7..145M}. 
which was characterised by a large increase in the mean brightness and
the appearance of strong and high-ionization emission lines
(HeII4686, [OIII]4959, 5007,  [NeIII]3869, etc.) and a prominent soft X-ray component 
\citep{2019MNRAS.489.2930Z}. 
Here, we analyse $UBV$ photometry obtained during and after the superactive state
and estimate key system parameters in terms of the colours and the optical luminosity
of the hot component.  

 \begin{figure*}     
  \vspace{10.0cm}   
  \includegraphics{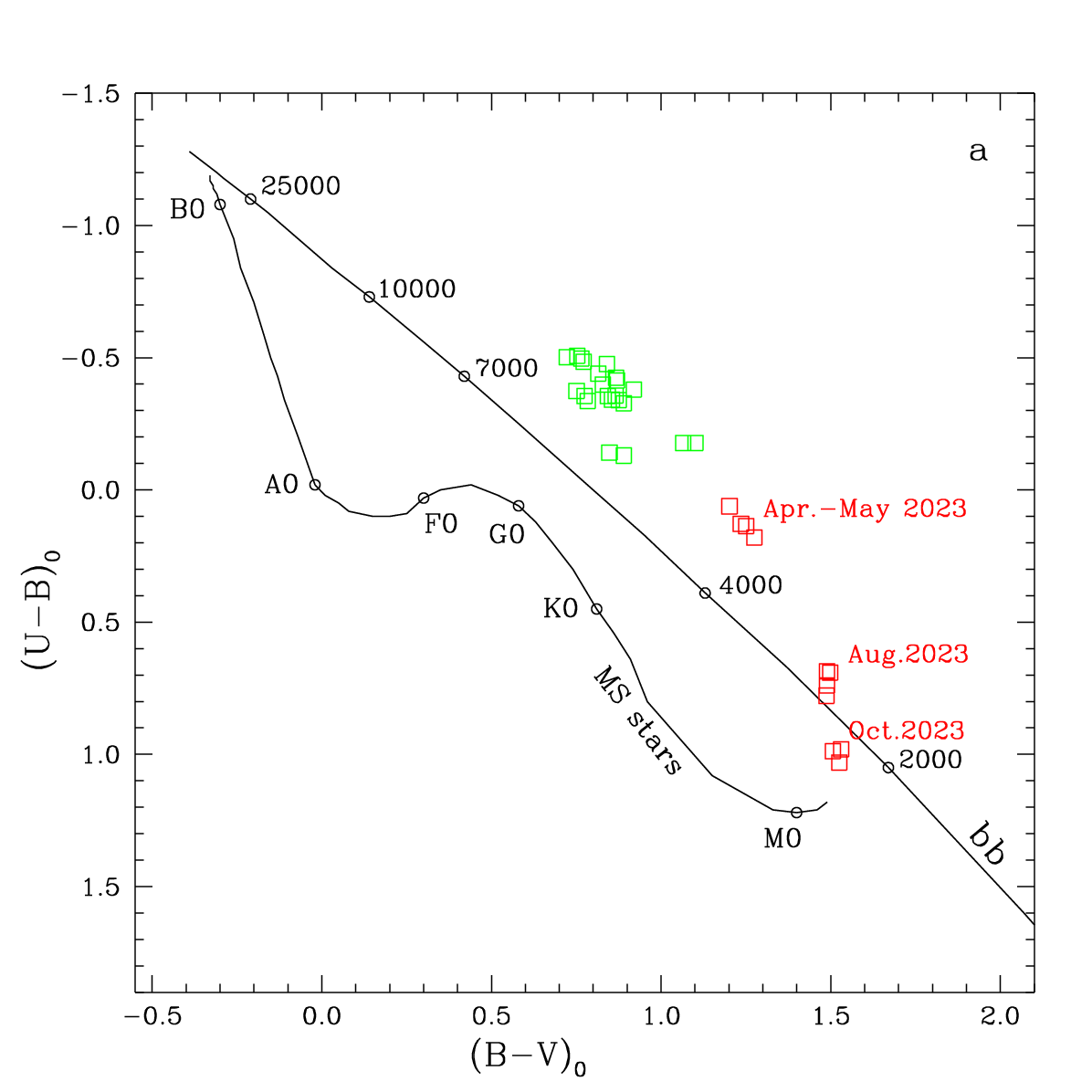} 
  \includegraphics{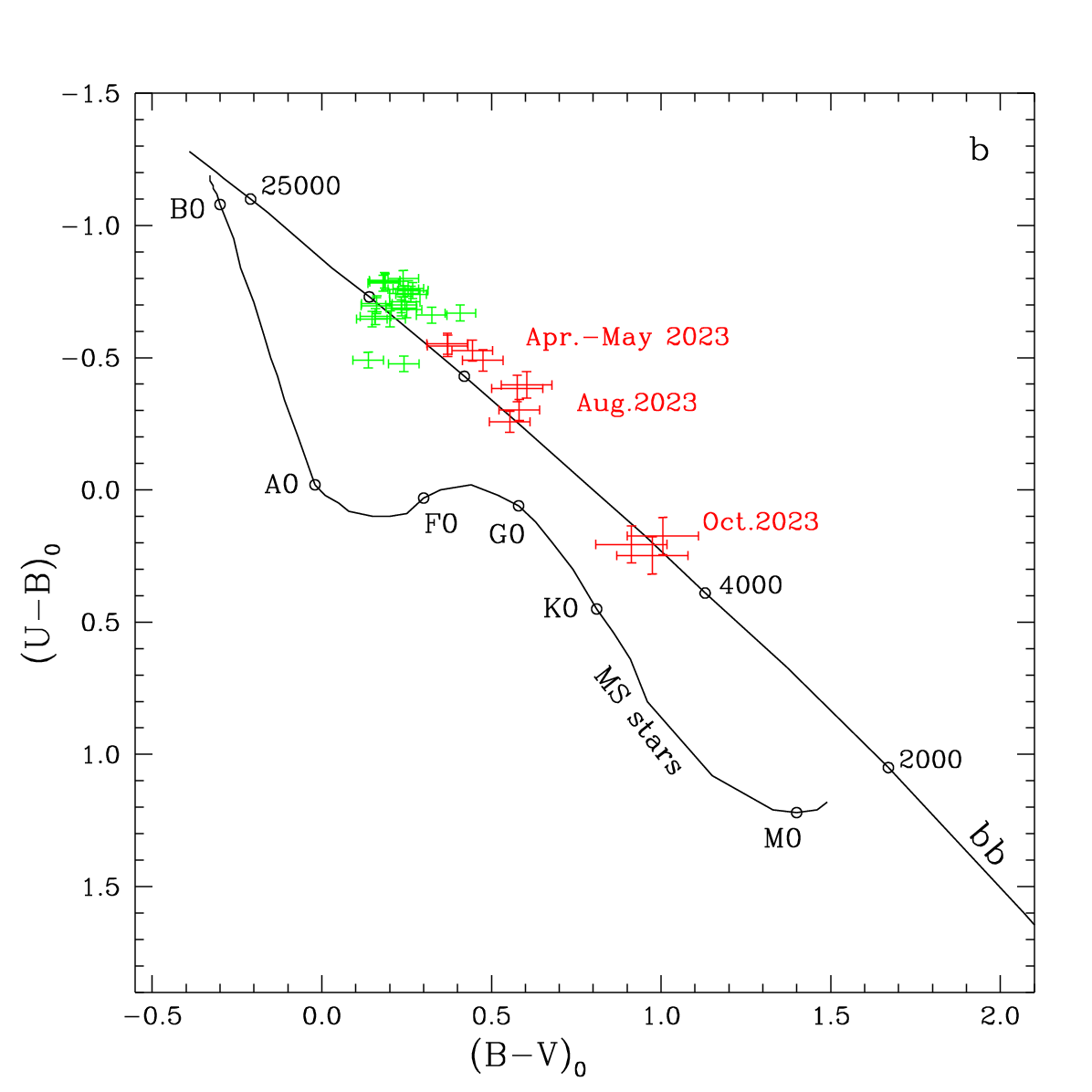}         
  \caption[]{Two-colour diagram $(U-B)_0$ versus $(B-V)_0$. 
           The left panel shows the dereddened colours of T~CrB.
           The right panel shows the hot component (red giant subtracted). 
           The green colours denote observations 
           obtained during the superactive state (2016-2022), and
           the red colours relate to observations made shortly afterwards           (April-October 2023). 
           More details can be found in Table~\ref{t.2}.   
           }
\label{f.2C}      
\end{figure*}        


\section{Observations}

$UBV$ photometry was obtained on 18 nights between 2016 and 2023. The observations were 
performed with the 50/70~cm Schmidt, the 1.5m,  and the 2.0m RCC telescopes  of the Rozhen  National Astronomical Observatory. 
Comparison stars from the list by \cite{2006A&A...458..339H}
were used. Some  of the data obtained were announced in The Astronomer's Telegrams
\citep{2016ATel.8675....1Z, 2023ATel16023....1M, 2023ATel16213....1Z}. \\



\begin{table*}
\caption{Dereddened colours and luminosity of the hot component.   }
\begin{center}
 \begin{tabular}{lcl ccc ccr cccll}
 \hline
 date         &   JD  & orb.phase & hot comp. & hot comp. & hot comp.  & $T_{eff}$ & $R_{eff}$    & $L_{opt}$    & $\dot M_a$ & \\
              &       &           &   $U_0$   & $(U-B)_0$ & $(B-V)_0$  &  [K]       &  [$R_\odot$] & $[L_\odot]$ & [$10^{-8} M_\odot$~yr$^{-1}$]& \\
  2016-02-07  &  57425.584   &    0.768   &  9.946   &  -0.800  &  0.240  &  10012 & 2.44 & 53.8  &  1.90  &   \\
  2016-04-01  &  57479.584   &    0.005   &  9.462   &  -0.647  &  0.201  &  9131  & 3.70 & 85.4  &  3.02  &   \\
  2016-04-03  &  57481.576   &    0.014   &  9.137   &  -0.740  &  0.263  &  9322  & 4.11 & 114.5 &  4.05  &   \\
  2017-02-23  &  57807.593   &    0.447   &  10.083  &  -0.491  &  0.137  &  8762  & 3.05 & 49.1  &  1.74  &   \\
  2017-02-24  &  57808.569   &    0.451   &  10.083  &  -0.477  &  0.242  &  8044  & 3.73 & 52.3  &  1.85  &   \\
  2017-03-28  &  57841.460   &    0.595   &  9.941   &  -0.697  &  0.161  &  9675  & 2.63 & 54.2  &  1.92  &   \\
  2017-03-28  &  57841.499   &    0.596   &  9.900   &  -0.704  &  0.162  &  9710  & 2.66 & 56.2  &  1.99  &   \\
  2017-04-26  &  57870.479   &    0.723   &  9.981   &  -0.711  &  0.245  &  9232  & 2.85 & 52.8  &  1.87  &   \\
  2017-04-27  &  57870.542   &    0.723   &  9.900   &  -0.743  &  0.245  &  9462  & 2.80 & 56.5  &  2.00  &   \\
  2018-01-24  &  58142.635   &    0.919   &  9.451   &  -0.793  &  0.185  &  10292 & 2.91 & 85.0  &  3.01  &   \\
  2018-01-24  &  58142.663   &    0.919   &  9.492   &  -0.787  &  0.187  &  10224 & 2.89 & 81.8  &  2.89  &   \\
  2018-01-24  &  58142.684   &    0.919   &  9.531   &  -0.782  &  0.181  &  10217 & 2.84 & 78.9  &  2.79  &   \\
  2018-04-14  &  58223.499   &    0.274   &  10.084  &  -0.759  &  0.240  &  9639  & 2.48 & 47.5  &  1.68  &   \\
  2018-04-15  &  58223.555   &    0.274   &  10.050  &  -0.762  &  0.255  &  9572  & 2.55 & 49.1  &  1.74  &   \\ 
  2018-07-06  &  58306.354   &    0.638   &  9.834   &  -0.656  &  0.158  &  9452  & 2.90 & 60.0  &  2.12  &   \\
  2018-07-06  &  58306.388   &    0.638   &  9.890   &  -0.646  &  0.148  &  9456  & 2.82 & 57.0  &  2.02  &   \\
  2020-04-16  &  58955.520   &    0.491   &  10.337  &  -0.661  &  0.324  &  8494  & 2.91 & 39.6  &  1.40  &   \\
  2020-04-16  &  58955.585   &    0.491   &  10.355  &  -0.669  &  0.408  &  8191  & 3.15 & 40.1  &  1.42  &   \\
  2021-01-20  &  59234.645   &    0.717   &  10.206  &  -0.682  &  0.250  &  9030  & 2.69 & 43.2  &  1.53  &   \\
  2021-01-20  &  59234.667   &    0.718   &  10.133  &  -0.700  &  0.234  &  9236  & 2.65 & 45.9  &  1.62  &   \\
  2021-01-20  &  59234.695   &    0.718   &  10.173  &  -0.688  &  0.235  &  9159  & 2.65 & 44.3  &  1.57  &   \\ 
  2022-07-23  &  59784.277   &    0.133   &  9.984   &  -0.754  &  0.268  &  9418  & 2.72 & 52.3  &  1.85  &   \\
  2023-04-28  &  60063.347   &    0.359   &  11.201  &  -0.490  &  0.475  &  7059  & 3.14 & 22.0  &  0.78  &   \\
  2023-04-28  &  60063.454   &    0.359   &  11.067  &  -0.527  &  0.444  &  7298  & 3.05 & 23.7  &  0.84  &   \\
  2023-05-24  &  60089.347   &    0.473   &  11.078  &  -0.553  &  0.370  &  7682  & 2.65 & 21.9  &  0.78  &   \\
  2023-05-24  &  60089.402   &    0.473   &  11.180  &  -0.545  &  0.371  &  7647  & 2.56 & 20.1  &  0.71  &   \\
  2023-08-11  &  60168.393   &    0.821   &  12.533  &  -0.384  &  0.576  &  6385  & 2.28 & 7.8   &  0.28  &   \\
  2023-08-11  &  60168.395   &    0.821   &  12.556  &  -0.397  &  0.604  &  6354  & 2.29 & 7.7   &  0.27  &   \\    
  2023-08-25  &  60182.292   &    0.882   &  12.526  &  -0.258  &  0.554  &  6103  & 2.64 & 8.7   &  0.31  &   \\
  2023-08-25  &  60182.361   &    0.882   &  12.466  &  -0.303  &  0.582  &  6143  & 2.66 & 9.0   &  0.32  &   \\
  2023-10-14  &  60232.218   &    0.101   &  12.661  &   0.206  &  0.913  &  4500  & $^a$ &       &        &   \\
  2023-10-19  &  60237.197   &    0.123   &  12.727  &   0.174  &  1.005  &  4400  & $^a$ &       &        &   \\
  2023-10-20  &  60238.201   &    0.127   &  12.819  &   0.248  &  0.974  &  4400  & $^a$ &       &        &   \\
 \hline                                                                                                        
 \end{tabular}                                                                                                           
 \end{center} 
  $^a$ the uncertainties are large due to the low brightness.                                                                                   
 \label{t.2}                                                                                                             
 \end{table*}                                                                                                     

\section{Results}
  
 


\subsection{System parameters}
Hereafter, for T~CrB,  we adopt an orbital period of   227.5687~d,  
T$_0$~(HJD)~2,447,918.62 \citep{2000AJ....119.1375F}, 
a mass of the white dwarf of $1.37 \pm 0.13$~$M_\odot$,
a mass of the red giant of  $1.12 \pm 0.23$~$M_\odot$, and an inclination of $i=67^o.5$ \citep{2004A&A...415..609S}. 
These are formal error bars, but the upper bound on the mass 
of the white dwarf is of course set by the Chandrasehkar Limit. 

Following Kepler's third law, the distance between the components is
$a=213$~$R_\odot$. We adopt a distance of $d=914$~pc \citep{2022MNRAS.517.6150S}, 
which is similar to the value  $d=890$~pc \citep{2021AJ....161..147B}
based on the $Gaia$~EDR3 \citep{2021A&A...649A...1G}. 
We also adopt interstellar extinction $E(B-V)=0.07$ \citep{2022NewA...9701859N}. 
This value is consistent with the Galactic dust reddening  maps 
by \cite{1998ApJ...500..525S} and \cite{2011ApJ...737..103S}, which give
$E(B-V) \le 0.071$ (calculated with the NASA/NED extinction calculator).

\subsection{UBV photometry}

The left panel of  Fig.~\ref{f.2C} shows a two-colour diagram,  
$(U-B)_0$ versus $(B-V)_0$, for T~CrB. Typical errors are $\pm 0.01$ mag. 
The right panel shows 
the colours of its hot component after the subtraction of the red giant contribution. 
The black solid lines are for main sequence stars \citep[taken from][]{1982sck} and for a black body.
The black body temperature is marked for some of the  points. 

From Fig.\ref{f.2C}a, it is apparent that during the superactive state (green symbols),
the colours of T~CrB are around $(U-B)_0 = -0.36 \pm 0.11$ and $(B-V)_0 = 0.85 \pm 0.09$. 
During 2023 (red symbols), T~CrB becomes gradually  redder  
$(U-B)_0 = 0.1$, $(B-V)_0 =1.2$ in April-May 2023,
$(U-B)_0 = 0.7$, $(B-V)_0 =1.5$ in August 2023, 
and $(U-B)_0 = 1.0$, $(B-V)_0 =1.5$ in October 2023.

\subsection{Cool component} 

\cite{1987AJ.....93..938K} 
analysed 
the prominent absorption features on the red spectra ($\lambda \lambda$ 5500 - 8600 \AA) 
of symbiotic stars and classified the cool component of T CrB as   M4~III.  
Using  eight-colour near-infrared (NIR) photometry,  \cite{2004AJ....128.2981K} 
find  the red giant is an M4.4~Ib - II.  
Based on NIR spectra in the range  7900-9300 \AA, 
\cite{1999A&AS..140...69Z} 
derived  M3.8~III. 
\cite{1999A&AS..137..473M}, 
also using NIR spectra (7000 - 10000 \AA), find a spectral type of M4.5. 

Using $d=914$~pc, $V=10.37$, and $E(B-V)=0.07$, 
we calculate an absolute $V$ band magnitude for the system in the low state $M_V \approx +0.4$, 
which is  one magnitude fainter than the expected $M_V \approx -0.6$ for a typical M4III star 
(e.g. Table 2 in \citealt{1981Ap&SS..80..353S}). 
Nevertheless, 
the value $M_V \approx +0.4$ is more appropriate 
for luminosity class III. It is not consistent with $-3 \ge M_V \ge -6$ 
expected for luminosity classes I-II, nor
with  $M_V \approx +10$ for luminosity class~V. 
For the cool component, we therefore adopt a spectral type of  M4~III, 
ellipsoidal variability in the form $V_0 = 10.16 - 0.2 \: \cos 4\pi \phi$,  
and colours  $U-B_0 = 1.65$, $B-V_0=1.63$, which 
are the colours of an M4.5~III giant \citep{1982sck}. 

\subsection{Accretion luminosity} 

From the $UBV$ data, we  calculate the  dereddened  
magnitude $U_0$, and the dereddened colours of the hot component.  
To estimate the luminosity of the hot component, we applied the following  procedure:
1. We correct the observed magnitudes for the interstellar extinction. 
2. We subtract the contribution of the  red giant 
using the calibrations for generic Bessell $UBV$  filters
given in \cite{2020sea..confE.182R}. 
3. We calculate  $U_0$, $(U-B)_0$ and $(B-V)_0$ of the hot component. 
4. Using $(U-B)_0$ and $(B-V)_0$ and the calibration for a black body 
(Table 18 in \citealt{1992msp..book.....S}), 
we calculate the effective temperature\footnote{The assumption of black body emission is a first
approximation, as the true spectrum is likely to be that of an accretion disc. 
However, the results shown in Fig. 1b indicate that black body 
emission is not an unreasonable assumption at this stage.}.
5. Using a distance of $d=914$~pc, and the dereddened magnitude $U_0$, 
we calculate the effective radius and the optical luminosity of the hot component.

Table~\ref{t.2} gives the date, Julian day, 
orbital phase, dereddened $U_0$ band magnitude of the hot component, dereddened colours 
$(U-B)_0$ and $(B-V)_0$ of the hot component, and its optical luminosity. 
The errors in the colours depend on the  subtraction of the red giant; in a bright state, they are  $\pm 0.03$ for $(U-B)_0$ and $\pm 0.06$ for $(B-V)_0$, 
and are two times larger in a low state. The typical error on 
the temperature is less than $\pm 700$~K.
The error on the estimated luminosity is 
of about $\pm 7$\%  in the bright state, and is two times larger in low state. 

\subsection{Temperature and luminosity  of the hot component}
From Fig.\ref{f.2C}b and Table~\ref{t.2}   it can be seen that, during the superactive state, 
the colours of the hot component cluster 
around $(U-B)_0 = -0.70 \pm 0.08$ and $(B-V)_0 = 0.23 \pm 0.06$. 
The average effective temperature 
during 2016-2021 is $9400 \pm 600$~K. 
During 2023, after the end of the superactive state, the hot component 
becomes gradually redder and its effective  temperature decreases from 9400~K
down to $\approx 5000$~K. 
We find that the optical luminosity of the hot component is $40-110~L_\odot$ during 
the  superactive state. This decreased 
to $20-25$~$L_\odot$ in April-May 2023, 
and to $8-9~L_\odot$ after the end of the superactive state in August 2023.

\subsection{Mass-accretion rate} 

The optical luminosity of the hot component is connected to  the mass-accretion rate: 
$L=0.5 \, G \, M_{wd} \, \dot M_a \,  \cos i  \;  R_{wd}^{-1}$ (supposing that the main source 
is the accretion disc), 
where $G$ is the gravitational constant, $\dot M_a$ is the mass-accretion rate, 
$R_{wd}$ is the white dwarf radius, and $i$ is the inclination. 
For the radius of the white dwarf, we adopt $R_{wd}=2018$~km, which we calculated using the Eggleton formula 
as given in \cite{1988ApJ...332..193V}. $\dot M_a$ calculated in this way is given in the last column of 
Table~\ref{t.2}.  

During the superactive state, we find an average of 
$\dot M_a \approx 2 \times  10^{-8}$~M$_\odot$~yr$^{-1}$
and a maximum of $4 \times 10^{-8}$~M$_\odot$~yr$^{-1}$
(the uncertainty is of about $\pm 30$~\%). 
However, if we assume the geometrical size of the boundary layer between 
the white dwarf and the accretion disc is approximately equal to two white dwarf radii 
\citep{1993A&A...275..219B}, then this value would be two times larger. 
If we assume a slightly lower mass of the white dwarf of
$M_{wd} = 1.2$~M$_\odot$ (Belczynski \& Mikolajewska 1998),
and a corresponding $R_{wd} =3859$~km, this latter value is two times larger.

This means that the total mass accreted during the superactive state 
from 2014 to 2023 is in the range $1.2 \times 10^{-7} -  5 \times 10^{-7}$~M$_\odot$, 
with the most likely value being  $1.7 \times 10^{-7}$~M$_\odot$.


\section{Discussion}
\cite{2020ApJ...902L..14L}  suggest that the white dwarfs in T~CrB and other 
symbiotic recurrent novae accumulate most of the ignition mass during sporadic high states. 
Hints of such episodes were also noticed by \cite{2011ApJ...737....7N}
in the  X-ray data of the symbiotic recurrent nova RS Oph.
In late 2014, T~CrB entered a superactive state that peaked in April 2016
and ended in mid-2023 \citep{2016NewA...47....7M, 2023RNAAS...7..145M}. 
\cite{2023ApJ...953L...7I} pointed out that this superactive state of T~CrB is similar to 
the superoutbursts observed in SU UMa-type dwarf novae. 

We find that the optical luminosity of the hot component was $50-110~L_\odot$ during 
the early stages of the  superactive state (February - April 2016) and stayed at a 
level $\approx 40-50~L_\odot$ until 2022. 
The luminosity decreased to $20-25$~$L_\odot$ in April-May 2023, 
and to $8-9~L_\odot$ after the end of the superactive state in August 2023. 
As the luminosity decreases, the hot component becomes redder and its effective temperature 
also decreases. 
There several possible drivers of the decrease in the optical brightness of T~CrB in 2023: 
(1) A decrease in the mass-accretion rate may have led to this decrease, because the accumulated material in the disc 
is exhausted and a new phase of accumulation begins.
(2) The red giant transferred a lot of material during the period 2014-2023, and then shrank and transferred less material via $L_1$. 
(3) \cite{2023ATel16107....1S} consider that this is a pre-outburst dip 
and is a signal that a nova eruption will be observed in the coming months.  
If it is a pre-outburst dip, it could 
be connected to the formation of a dense,  near-to-critical mass envelope 
around the white dwarf \citep{2023arXiv230804104Z}.

\cite{2008ASPC..401...73W} discussed the possibility that the accretion disc in 
a symbiotic recurrent nova acts as a reservoir 
of mass containing up to $10^{-6}$~M$_\odot$.
The disc is large, with the mass required for the TNR being accreted during 
a few large and bright disc outbursts \citep[for more details, see][sect.~3.1]{2008ASPC..401...73W}.
In such a scenario, during the disc outburst phases, the accretion disc is hot 
and the accretion rate is high.
Our results for the behaviour of T~CrB during 2016-2023 (see Fig.1a,b and Table~1) 
are in good agreement with these considerations. If a nova outburst happens 
over the following 12 months, this would be an indication that one superactive state 
(disc superoutburst of SU~UMa type) 
can be sufficient to cause
a TNR.


The total mass accreted during the superactive state 
from 2014 to 2022 is $\approx 2 \times 10^{-7}$~M$_\odot$, which  
is a considerable fraction ($\sim 30 \% $) of the amount that needs to be accumulated 
for a TNR 
\citep[$5 \times 10^{-7} - 1.6 \times 10^{-6}$~M$_\odot$;][]{2020A&A...634A...5J, 2018ApJ...860..110S}.
Whether or not the  material accumulated on the white dwarf is sufficient to cause
a TNR, resulting in a nova eruption, is an open question, which may well be answered in the coming year.

\section{Conclusions}
We analysed $UBV$ photometry of the recurrent nova T~CrB obtained during 2016-2023. 
After subtraction of the contribution of the red giant, 
we find average dereddened colours of the hot component of 
$(U-B)_0 = -0.70 \pm 0.08$ and $(B-V)_0 = 0.23 \pm 0.06$ during the period 
2016-2022.  
We estimate the optical luminosity of the hot component to be
$40-110~L_\odot$ during the superactive state. 
After the end of the superactive state, the hot component became 
redder and its luminosity decreased to $20-25$~$L_\odot$ in April-May 2023, 
and to $8-9~L_\odot$ in August 2023.  
The total mass accreted during the superactive state 
from 2014 to 2023 is $\sim 2 \times 10^{-7}$~M$_\odot$, 
which is $\sim 30 \%$ of the total mass needed
for a nova outburst powered by a TNR.  



\begin{acknowledgements} 
The research infrastructure is supported by the Ministry of Education and Science of Bulgaria 
(Bulgarian National Roadmap for Research Infrastructure) and by the  Bulgarian National Science Fund.
DM acknowledges support by project RD-08-157/2023 from Shumen University Science Fund. \\

\end{acknowledgements}

%
%

\bibliographystyle{aa}
\bibliography{ref3.bib}

\newpage 


\begin{table*}
\caption{UBV observations of T CrB. This table contains the photometry used to calculate the values 
in Table~\ref{t.2}. The observations of the intra-night variability (flickering)
are available on zenodo.org/records/10283430.  }
\begin{center}
 \begin{tabular}{lcl ccc ccr cccll}
 \hline
yyyy-mm-dd &       hh:mm	 &    U     &    B     &    V       &   \\
\\
2016-02-07 &       00:49-03:53   &  10.262  &  10.738  &  9.8980    &   \\  	  
2016-04-01 &       01:16-03:16   &  9.781   &  10.155  &  9.4040    &   \\
2016-04-03 &       00:28-02:49   &  9.465   &  9.967   &  9.2450    &   \\
2017-02-23 &       02:02-02:29   &  10.380  &  10.521  &  9.6731    &   \\
2017-02-24 &       01:33-01:51   &  10.380  &  10.510  &  9.6200    &   \\
2017-03-28 &       22:06-23:45   &  10.248  &  10.589  &  9.7339    &   \\
2017-03-28 &       23:46-01:07   &  10.208  &  10.564  &  9.7203    &   \\
2017-04-26 &       23:12-23:59   &  10.296  &  10.695  &  9.8662    &   \\
2017-04-27 &       00:01-01:23   &  10.217  &  10.657  &  9.8424    &   \\
2018-01-24 &       03:00-03:35   &  9.773   &  10.281  &  9.5275    &   \\
2018-01-24 &       03:37-04:18   &  9.813   &  10.309  &  9.5438    &   \\
2018-01-24 &       04:19-04:31   &  9.851   &  10.335  &  9.5629    &   \\
2018-04-14 &       23:38-00:45   &  10.395  &  10.809  &  9.9394    &   \\ 
2018-04-15 &       00:47-01:40   &  10.362  &  10.786  &  9.9187    &   \\
2018-07-06 &       20:09-20:52   &  10.149  &  10.504  &  9.7299    &   \\
2018-07-06 &       20:53-21:30   &  10.203  &  10.540  &  9.7564    &   \\
2020-04-16 &       00:58-01:30   &  10.617  &  10.795  &  9.7300    &   \\
2020-04-16 &       01:32-02:11   &  10.634  &  10.811  &  9.7100    &   \\
2021-01-20 &       03:14-03:43   &  10.5126 &  10.8406 &  9.9503    &   \\
2021-01-20 &       03:44-04:18   &  10.4424 &  10.8002 &  9.9337    &   \\ 
2021-01-20 &       04:19-04:54   &  10.4806 &  10.8206 &  9.9450    &   \\
2022-07-23 &       18:39-18:41   &  10.29   &  10.67   &  9.75      &   \\
2023-04-28 &       19:49-21:35   &  11.4275 &  11.2986 &  10.0631   &   \\
2023-04-28 &       21:35-23:49   &  11.3077 &  11.2460 &  10.0442   &   \\  	  
2023-05-24 &       20:10-20:59   &  11.2926 &  11.1563 &  9.9047    &   \\  	  
2023-05-24 &       21:00-21:51   &  11.3815 &  11.2014 &  9.9257    &   \\  	  
2023-08-11 &       ...21:27	 &  12.531  &  11.846  &  10.357    &   \\  	  
2023-08-11 &       ...21:30	 &  12.547  &  11.856  &  10.358    &   \\  	  
2023-08-25 &       18:47-19:24   &  12.484  &  11.7058 &  10.2181   &   \\  	  
2023-08-25 &       19:26-20:50   &  12.442  &  11.7018 &  10.2127   &   \\  	  
2023-10-14 &       ...17:15	 &  12.562  &  11.574  &  10.068    &   \\
2023-10-19 &       ...16:44	 &  12.624  &  11.643  &  10.112    &   \\
2023-10-20 &     ...16:50        &  12.687  &  11.656  &  10.131    &   \\ 
2023-10-23 &     ...16:50        &  12.785  &  11.707  &  10.172    &   \\ 
2023-10-23 &     ...17:25        &  12.767  &  11.699  &  10.158    &   \\ 
 \hline                            									     
 \end{tabular}                                                                                                           
 \end{center} 
 \label{t.4}                                                                                                             
 \end{table*}                                                                                                     

\end{document}